\documentclass [12pt]{article}
\usepackage {graphicx}

\begin{document}
\begin{center}
\textbf{Direct reconstruction of the effective atomic number of materials by
the method of multi-energy radiography}
\end{center}

\bigskip

\begin{center}
\textbf{Sergey V. Naydenov and Vladimir D. Ryzhikov}
\end{center}

\begin{center}
\textbf{Institute of Single Crystals}
\end{center}

\begin{center}
\textbf{60 Lenin Avenue, Kharkov, 61001 Ukraine}
\end{center}

\bigskip

\begin{center}
\textbf{Craig F. Smith}
\end{center}

\begin{center}
\textbf{Lawrence Livermore National Laboratory}
\end{center}

\begin{center}
\textbf{PO Box 808, Livermore, CA 94556 USA}
\end{center}

\bigskip

A direct method is proposed for reconstruction of the effective atomic
number by means of multi-energy radiography of the material. The accuracy of
the method is up to 95\% . Advantages over conventional radiographic
methods, which ensure accuracy of just about 50\% , are discussed. A
physical model has been constructed, and general expressions have been
obtained for description of the effective atomic number in a two-energy
monitoring scheme. A universal dependence has been predicted for the
effective atomic number as a function of relative (two-energy) radiographic
reflex. The established theoretical law is confirmed by the experimental
data presented. The proposed development can find multiple applications in
non-destructive testing and related fields, including those in the civil
sphere as well as anti-terrorist activities.

\bigskip

07.85.-m ,78.20.Bh ,81.70.-q, 81.70.Fy ,87.59.Hp

\newpage
\begin{center}
\textbf{1. Introduction}
\end{center}

Non-destructive testing (NDT) of different unknown materials,
objects and media, including complex micro- and macro-structures,
is an important field in modern applied physics \cite{15th:2000}.
Physical methods of NDT find multiple applications in a wide
variety of technical areas, including quality control of
industrial products, inspection of welded joints and connections,
and confirmation of building construction, pipelines, etc. An
important role is also played by NDT in radioactive materials
control \cite{Robert:2003}, security protection in aviation
\cite{NRC:1996}, \cite{Grodzins:1991}, railway and automobile
transport, and customs inspection of goods and luggage. In
addition, broad applications related to the use of new NDT
technologies in medicine \cite{Barnes:1985}, \cite{Feldkamp:1989},
\cite{Inanc:1998}, \cite{Marziani:2002}, including separate
diagnostics of tissues and organs, can be expected. Among the
promising directions of NDT, one should note digital (computer)
radiography. In it, digital reconstruction of an image of an
object examined by X-ray or gamma-radiation is prepared based on
the data collected in real-time detection. The most widely used
radiation detectors for such applications are solid-state 1D and
2D scintillation systems \cite{Harrison:1991},
\cite{Rapiscan:2002}, \cite{Heimann:2002}, \cite{YXLON:2002},
\cite{Poliscan:1}.

The main task of digital radiography, as well as of other NDT
methods, is reconstruction of the physical parameters that
determine technical characteristics and properties of the
controlled objects being examined. This is traditionally carried
out by reconstruction of the spatial configuration, observing
small details and defects of the image, and determining
constitutive components and dimensions (thickness). In recent
times, strong interest has been emerging in the development of
direct methods for reconstruction of the overall composition of
the object, i.e., quantitative determination of the effective
atomic number, density, chemical composition, etc. To a large
extent, this is due to improvements in technical capabilities for
detection and image processing and is driven by the desire for
additional detail in the image resulting from the collected data.
Often it is not sufficient to know just the geometrical structure
of an object to ensure its complete identification for
decision-making purposes. For example, fully reliable detection of
explosives is not possible without discrimination of materials by
their effective atomic number (organics and inorganics), density,
content of predominant chemical elements, etc. The importance of
such enhanced monitoring cannot be overestimated. It is vitally
necessary not only to provide enhanced tools for scientific and
technological investigations, but to meet current needs for
improved protection against terrorist threats to the safety and
health of civil populations. An important milestone on the way to
solving this problem is the implementation of two-, and in
general, of multi-energy radiography \cite{Alvarez:1976},
\cite{Coenen:1994}, \cite{Rizescu:2001}, \cite{Naydenov:1}.

\bigskip

\begin{center}
\textbf{2. Effective atomic number}
\end{center}

Among the parameters determining the constitutive structure of an unknown
object or material, one should especially note the effective atomic number
$Z_{{\rm e}{\rm f}{\rm f}} $. In fact, this measure can provide an initial
estimation of the chemical composition of the material. A large $Z_{{\rm
e}{\rm f}{\rm f}} $ generally corresponds to inorganic compounds and metals,
while a small $Z_{{\rm e}{\rm f}{\rm f}} \le 10$ is an indicator of organic
substances. For many applications, e.g., radioisotope monitoring,
cross-section studies of absorption, scattering and attenuation of
electromagnetic radiation, testing of multi-component, heterogeneous and
composite materials, etc., this parameter is of principal significance. It
is very important to obtain its quantitative evaluation from the data
provided by radiographic measurements. With the aim of making this possible,
in the present work we have developed a physical model for two-energy
radiography as a particular case of the multi-energy radiography. We show
that a multiplicity of $M = 2$ is sufficient for direct determination of
$Z_{{\rm e}{\rm f}{\rm f}} $. General theoretical expressions are provided
for determining the effective atomic number with respect to the photo and
pair generation effects.

The method proposed here for determination of $Z_{{\rm e}{\rm f}{\rm f}} $
is substantially different from the approaches used previously. In these
procedures, the effective atomic number is comprised of a linear
superposition of several reference values, i.e., it is expressed in the form
of $Z_{{\rm e}{\rm f}{\rm f}} = {\sum\nolimits_{k = 1}^{N} {p_{k} Z_{k}}
}$, where ${\sum\nolimits_{k} {p_{k}} }  = 1$, and $N = 2,3,...$ corresponds
to the multiplicity of the radiography used. Such ``synthesis'' of $Z_{{\rm
e}{\rm f}{\rm f}} $ is quite similar to the formation of light of arbitrary
color from several basic colors (Fig.1). Weights $p_{k} $, or the relative
content of the components (light$Z_{L} $, heavy$Z_{H} $, intermediate$Z_{M}
$), are determined from the data of radiographic measurements.
Theoretically, this corresponds to the solution of systems of linear
equations of the form $R_{i} = {\sum\nolimits_{k} {\mu _{ik} {\kern 1pt}
d_{k}} } $ (where $d_{k} $ is the layer thickness of the $k$-th element
phantom in the studied object), but with attenuation coefficients $\mu _{ik}
= \mu \,\left( {E_{i} ;Z_{k} ,\rho _{k}}  \right)$ that are \textit{a priori} pre-set at
several fixed energy values $E_{i} $ for the basic elements, each with its
selected atomic number $Z_{k} $ and density $\rho _{k} $. The values of
$p_{k} = {{d_{k}}  \mathord{\left/ {\vphantom {{d_{k}}  {{\sum\nolimits_{k}
{d_{k}} } }}} \right. \kern-\nulldelimiterspace} {{\sum\nolimits_{k} {d_{k}
}} }}$ depend not only on reflexes $R_{i} $, but also on the basis chosen.
Therefore, such radiography is, in fact, ``indirect''. Erroneous or
inaccurate choice of the basis leads to mistakes (artifacts) in
determination of $Z_{{\rm e}{\rm f}{\rm f}} $. For example, a sample with
low content of a heavy element (low $Z_{H} $) can be mistakenly identified
as a sample with a thick layer of a low $Z_{L} $ element. It is clear that,
expanding the basis and increasing the multiplicity, one can substantially
reduce the errors. However, even then, errors can be as high as tens of
percent. The accuracy of traditional methods in 2- and 3-energy radiography
allows distinction of materials with difference in $Z_{{\rm e}{\rm f}{\rm
f}} $ by about 50\% . So, it is possible to distinguish iron alloys ($Z
\approx 26$) from wood ($Z \approx 6$) and even from light aluminum alloys
($Z \approx 13$), but it is not possible to reliably discern the difference
between water ($Z \approx 8$) and organic materials ($Z \approx 6 - 7$),
iron and many other metals, calcium minerals ($Z \approx 17$) and sand ($Z
\approx 12$), etc. To distinguish among these materials requires accuracy
that should be higher by an order of magnitude (i.e., not worse than 95\% ).

For a chemical compound with the known formula of the form ${\left[ {X}
\right]} = {\left[ {{}_{Z_{{\rm e}{\rm f}{\rm f}}} X^{A_{{\rm e}{\rm f}{\rm
f}}} } \right]} = {\left\{ {{\left[ {{}_{Z_{i}} X^{A_{i}} } \right]}_{n_{i}
}}  \right\}}_{i = 1}^{N} $ (here $n_{i} $ is the number of atoms of the
$i$-th kind with atomic number $Z_{i} $ and atomic mass $A_{i} $ in the
complex molecule ${\left[ {X} \right]}$; $N$ is the full number of simple
components), $Z_{{\rm e}{\rm f}{\rm f}} $ and $A_{{\rm e}{\rm f}{\rm f}} $
are determined from the expressions

\begin{equation}
\label{eq1}
Z_{{\rm e}{\rm f}{\rm f}} = {\left[ {{{{\sum\limits_{i = 1}^{N} {n_{i} A_{i}
Z_{i}^{p + 1}} } } \mathord{\left/ {\vphantom {{{\sum\limits_{i = 1}^{N}
{n_{i} A_{i} Z_{i}^{p + 1}} } } {{\sum\limits_{i = 1}^{N} {n_{i} A_{i} Z_{i}
}} }}} \right. \kern-\nulldelimiterspace} {{\sum\limits_{i = 1}^{N} {n_{i}
A_{i} Z_{i}} } }}} \right]}^{{{1} \mathord{\left/ {\vphantom {{1} {p}}}
\right. \kern-\nulldelimiterspace} {p}}}\;{\rm ;}
\quad
A_{{\rm e}{\rm f}{\rm f}} = {{{\sum\limits_{i = 1}^{N} {n_{i} A_{i}} } }
\mathord{\left/ {\vphantom {{{\sum\limits_{i = 1}^{N} {n_{i} A_{i}} } }
{{\sum\limits_{i = 1}^{N} {n_{i}} } }}} \right. \kern-\nulldelimiterspace}
{{\sum\limits_{i = 1}^{N} {n_{i}} } }}{\rm ,}
\end{equation}

\noindent
where $p$ is an index depending upon the chosen absorption mechanism of
gamma-quanta. The expression (\ref{eq1}) for determination of $Z_{{\rm e}{\rm f}{\rm
f}} $ is derived from the balance of the absorbed energy per each atomic
electron of the substance. Absorption over all possible physical channels
and by all simple components of the complex compound is accounted for.
Parameter $p$ is related to the characteristic dependence of the electron
coefficient of linear absorption, $\mu _{e} = {{\mu _{m}}  \mathord{\left/
{\vphantom {{\mu _{m}}  {Z}}} \right. \kern-\nulldelimiterspace} {Z}} =
{{\mu}  \mathord{\left/ {\vphantom {{\mu}  {\rho Z}}} \right.
\kern-\nulldelimiterspace} {\rho Z}}$ (here $\mu _{m} $ is the mass
attenuation coefficient), on the atomic numbers of simple constituents of
the substance studied. The choice of $p$ values corresponds to the
predominant absorption channels at specified radiation energies, i.e. $\mu
_{e} \propto Z_{{\rm e}{\rm f}{\rm f}}^{p} $. In the photo effect theory,
$\left( {\mu _{e}}  \right)_{{\rm p}{\rm h}{\rm o}{\rm t}{\rm o}} \propto
Z^{3}$. In the pair formation theory $\left( {\mu _{e}}  \right)_{{\rm
p}{\rm a}{\rm i}{\rm r}{\rm s}} \propto Z$. For the Compton scattering
weakly depending upon atomic properties of the material $\left( {\mu _{m}}
\right)_{{\rm c}{\rm o}{\rm m}{\rm p}{\rm t}} \propto Z^{0} \Rightarrow
\left( {\mu _{e}}  \right)_{{\rm c}{\rm o}{\rm m}{\rm p}{\rm t}} \propto Z^{
- 1}$. In the experiments, such absorption character is also well confirmed.
Besides this, the parameter $p$ can be considered as a fitting parameter
giving the best approximation of the absorption cross-section in a specified
energy range and for a specified class of substances. E.g., in the photo
effect energy range intermediate values $p = 2\div 4$ are used. We use for
the photo effect $p = 3$, while for the pair formation effect, $p = 1$, i.e.

\begin{equation}
\label{eq2}
Z_{{\rm p}{\rm h}{\rm o}{\rm t}{\rm o}} = {\left[ {{{{\sum\limits_{i =
1}^{N} {n_{i} A_{i} Z_{i}^{4}} } } \mathord{\left/ {\vphantom
{{{\sum\limits_{i = 1}^{N} {n_{i} A_{i} Z_{i}^{4}} } } {{\sum\limits_{i =
1}^{N} {n_{i} A_{i} Z_{i}} } }}} \right. \kern-\nulldelimiterspace}
{{\sum\limits_{i = 1}^{N} {n_{i} A_{i} Z_{i}} } }}} \right]}^{{{1}
\mathord{\left/ {\vphantom {{1} {3}}} \right. \kern-\nulldelimiterspace}
{3}}}\;{\rm ;}
\quad
Z_{{\rm p}{\rm a}{\rm i}{\rm r}{\rm s}} = {\left[ {{{{\sum\limits_{i =
1}^{N} {n_{i} A_{i} Z_{i}^{2}} } } \mathord{\left/ {\vphantom
{{{\sum\limits_{i = 1}^{N} {n_{i} A_{i} Z_{i}^{2}} } } {{\sum\limits_{i =
1}^{N} {n_{i} A_{i} Z_{i}} } }}} \right. \kern-\nulldelimiterspace}
{{\sum\limits_{i = 1}^{N} {n_{i} A_{i} Z_{i}} } }}} \right]}\;{\rm .}
\end{equation}

For the Compton scattering, which normally accompanies one of the former
mechanisms, it should be assumed by definition that $p = - 1$ and

\begin{equation}
\label{eq3}
Z_{{\rm c}{\rm o}{\rm m}{\rm p}{\rm t}} = {{{\sum\limits_{i = 1}^{N} {n_{i}
A_{i} Z_{i}} } } \mathord{\left/ {\vphantom {{{\sum\limits_{i = 1}^{N}
{n_{i} A_{i} Z_{i}} } } {{\sum\limits_{i = 1}^{N} {n_{i} A_{i}} } }}}
\right. \kern-\nulldelimiterspace} {{\sum\limits_{i = 1}^{N} {n_{i} A_{i}}
}}}\quad {\rm ,}
\end{equation}

\noindent
as the Compton effect cross-section does not depend upon properties of the
material, but only upon its average electron density. The molar mass $M_{X}
= {\sum\nolimits_{i} {n_{i} A_{i}} } $. In practice, relative concentrations
$c_{i} $ of the simple components of the material are often also known.
Then, as $c_{i} = {{n_{i}}  \mathord{\left/ {\vphantom {{n_{i}}
{{\sum\nolimits_{i} {n_{i}} } }}} \right. \kern-\nulldelimiterspace}
{{\sum\nolimits_{i} {n_{i}} } }}$ (where $0 \le c_{i} \le 1$), instead of
(\ref{eq1})-(\ref{eq2}) we obtain

\begin{equation}
\label{eq4}
Z_{{\rm e}{\rm f}{\rm f}} = {\left[ {{{{\sum\limits_{i = 1}^{N} {c_{i} A_{i}
Z_{i}^{p + 1}} } } \mathord{\left/ {\vphantom {{{\sum\limits_{i = 1}^{N}
{c_{i} A_{i} Z_{i}^{p + 1}} } } {{\sum\limits_{i = 1}^{N} {c_{i} A_{i} Z_{i}
}} }}} \right. \kern-\nulldelimiterspace} {{\sum\limits_{i = 1}^{N} {c_{i}
A_{i} Z_{i}} } }}} \right]}^{{{1} \mathord{\left/ {\vphantom {{1} {p}}}
\right. \kern-\nulldelimiterspace} {p}}}\;{\rm ;}
\quad
A_{{\rm e}{\rm f}{\rm f}} = {\sum\limits_{i = 1}^{N} {c_{i} A_{i}} } \quad
{\rm .}
\end{equation}

In Table 1, data are presented on the effective atomic number of selected
substances that comprise many materials in practical use. The data were
calculated using formulas (\ref{eq1})-(\ref{eq4}). It should be noted that the effective
atomic number is dependent upon the energy range of the ionizing radiation
used. Its value corresponds to the predominant absorption channel. Using
elementary inequalities from expressions (\ref{eq1})-(\ref{eq3}), one can easily obtain the
general relationship valid for chemical compounds:

\begin{equation}
\label{eq5}
Z_{{\rm c}{\rm o}{\rm m}{\rm p}{\rm t}} \le Z_{{\rm p}{\rm a}{\rm i}{\rm
r}{\rm s}} \le Z_{{\rm p}{\rm h}{\rm o}{\rm t}{\rm o}} \quad {\rm ,}
\end{equation}

\noindent
which is substantiated by the data of Table 1. For homogeneous mixtures
(solid, liquid, gaseous), including alloys, these inequalities (\ref{eq5}) can be
violated at certain concentration ratios of the mixture components. This
feature (inversion) can be used, for example, in identification of
substitution alloys or composites.

\begin{center}
\textbf{3. Theory and physical model of two-radiography}
\end{center}

Let us consider a simple basic model of two-energy radiography used for
direct qualitative determination (monitoring) of the effective atomic number
of a material. A general scheme of such radiography is presented in Fig.~1.
X-ray and gamma-radiation is attenuated exponentially with linear
coefficient $\mu = \mu \left( {E;\rho ;{\left[ {X} \right]}} \right)$. The
latter depends upon the radiation energy $E$, the density of the material
$\rho $, and its chemical (atomic) composition. $Z_{{\rm e}{\rm f}{\rm f}} $
is considered as a direct characteristic of the atomic composition. For
simplicity, we assume that: 1) the radiation is monochromatic at the two
fixed energies $E_{1} $ and $E_{2} $; 2) its spectrum is not changed when
passing through the object, and 3) scattered radiation can be neglected. By
the appropriate choice of energy filters, or by using radioactive isotopes
as radiation sources and subsequent collimation of the radiation in a
sufficiently narrow beam, rather good approximation of these conditions can
be realized in the experiment. Corrections for these factors can be also
accounted for theoretically. The non-monochromatic character of the emitted
and detected radiation, as well as its accumulation due to scattering inside
the object (and/or detector), generally lower the monitoring efficiency and
relative sensitivity.

Let us write down the (digitized) signal as recorded by detectors in the
form

\begin{equation}
\label{eq6}
V_{i} = V\left( {E_{i}}  \right) = \,V_{0} \left( {E_{i}}  \right)\exp
{\left[ { - \mu \left( {E_{i}}  \right)d} \right]}\quad {\rm ;}
\quad
i = 1,2 \quad {\rm ,}
\end{equation}

\noindent
where $V_{0} $ is the background signal (without an object); $d$is thickness
of local cross-section of the object in the direction of ray propagation.
One should note that the value of $V_{0i} = V_{0} \left( {E_{i}}  \right)$
depends upon full conversion efficiency of the system (i.e., the ratio of
the energy of the useful electron signal to the energy of the initial
photon) and upon the radiation source power. Let us separate the dependence
on the effective atomic number $Z_{{\rm e}{\rm f}{\rm f}} \equiv Z$ in the
attenuation coefficient

\begin{equation}
\label{eq7}
\mu \left( {E_{i}}  \right) = {\left[ {\tau \left( {E_{i}}  \right)Z^{4} +
\sigma \left( {E_{i}}  \right)Z + \chi \left( {E_{i}}  \right)Z^{2}}
\right]}\rho \quad {\rm ,}
\end{equation}

\noindent
where the functions $\tau \left( {E} \right),\;\sigma \left( {E}
\right),\,\chi \left( {E} \right)$ define the energy dependence (assumed to
be universal) of the actual absorption cross-sections for the photo effect,
Compton effect, and pair formation effect, respectively. It should be noted
that theoretical determination of $Z_{{\rm e}{\rm f}{\rm f}} $ for a complex
material using formulas (\ref{eq1})-(\ref{eq4}) is based just on the absorption structure as
given by (\ref{eq7}). Let us define the reflex as $R_{i} = R\left( {E_{i}}  \right)
= \ln {\left[ {{{V_{0i}}  \mathord{\left/ {\vphantom {{V_{0i}}  {V_{i}} }}
\right. \kern-\nulldelimiterspace} {V_{i}} }} \right]}$ and go from the
system of equation (\ref{eq6}) to the following linear equations:

\begin{equation}
\label{eq8}
{{R_{i}}  \mathord{\left/ {\vphantom {{R_{i}}  {\left( {\rho d} \right)}}}
\right. \kern-\nulldelimiterspace} {\left( {\rho d} \right)}} = \mu _{mi}
\equiv {\left[ {\alpha _{i} Z^{p} + \beta _{i}}  \right]}Z\quad {\rm .}
\end{equation}

Here, we have introduced the mass attenuation coefficient $\mu _{m} = {{\mu
} \mathord{\left/ {\vphantom {{\mu}  {\rho} }} \right.
\kern-\nulldelimiterspace} {\rho} }$ and the monitoring constants $\alpha
,\,\beta $, which depend only upon the radiation energy, but not on the
properties of the tested material. It is taken into account that the normal
energy ranges of the photo effect ($E \le 0.6MeV$) and the pair formation
effect ($E \ge 1.2MeV$), are, as a rule, sufficiently far from each other on
the energy scale. Hereafter, we will consider just the two-channel
absorption mechanism, i.e., either of the photo effect/Compton effect type
($p = 3$) or pair generation/Compton type ($p = 1$). Monitoring using only
the inverse Compton scattering corresponds to $p = - 1$. In all cases, it is
the relative reflex of 2-radiography that plays the principal role:

\begin{equation}
\label{eq9}
X = {{R_{{\kern 1pt} 1}}  \mathord{\left/ {\vphantom {{R_{{\kern 1pt} 1}}
{R_{{\kern 1pt} 2}} }} \right. \kern-\nulldelimiterspace} {R_{{\kern 1pt} 2}
}} \equiv {\frac{{\ln {\left[ {{{V_{0} \left( {E_{1}}  \right)}
\mathord{\left/ {\vphantom {{V_{0} \left( {E_{1}}  \right)} {{\kern 1pt}
V\left( {E_{1}}  \right)}}} \right. \kern-\nulldelimiterspace} {{\kern 1pt}
V\left( {E_{1}}  \right)}}} \right]}}}{{\ln {\left[ {{{V_{0} \left( {E_{2}}
\right)} \mathord{\left/ {\vphantom {{V_{0} \left( {E_{2}}  \right)} {{\kern
1pt} V\left( {E_{2}}  \right)}}} \right. \kern-\nulldelimiterspace} {{\kern
1pt} V\left( {E_{2}}  \right)}}} \right]}}}}\quad {\rm .}
\end{equation}

As it follows from equations (\ref{eq8}), this value is related to the relative
attenuation coefficient, $X \equiv {{\mu _{m} \left( {E_{1}}  \right)}
\mathord{\left/ {\vphantom {{\mu _{m} \left( {E_{1}}  \right)} {\mu _{m}} }}
\right. \kern-\nulldelimiterspace} {\mu _{m}} }\left( {E_{2}}  \right) =
{{\mu \left( {E_{1}}  \right)} \mathord{\left/ {\vphantom {{\mu \left(
{E_{1}}  \right)} {\mu} }} \right. \kern-\nulldelimiterspace} {\mu} }\left(
{E_{2}}  \right)$. Therefore $X$ does not depend upon the geometry
(thickness) or the density of the material nor does it depend on its other
physico-chemical properties, except its effective atomic number. In the
energy range where total absorption coefficient varies monotonically, we
have $0 < X < 1$, if $\mu \left( {E_{1}}  \right) < \mu \left( {E_{2}}
\right)$, or $X > 1$, if $\mu \left( {E_{1}}  \right) > \mu \left( {E_{2}}
\right)$.

Solving the system of equations (\ref{eq8}) with respect to unknown variables $Z$
and $\theta \equiv \left( {\rho \,d} \right)$ leads, after reconstruction of
the monitoring constants, to the expression

\begin{equation}
\label{eq10}
Z^{p} = {{{\left[ {\left( {Z_{1}^{p + 1} \theta _{1}}  \right)r_{1} - \left(
{Z_{2}^{p + 1} \theta _{2}}  \right)r_{2}}  \right]}} \mathord{\left/
{\vphantom {{{\left[ {\left( {Z_{1}^{p + 1} \theta _{1}}  \right)r_{1} -
\left( {Z_{2}^{p + 1} \theta _{2}}  \right)r_{2}}  \right]}} {{\left[
{\left( {Z_{1} \;\theta _{1}}  \right)\;r_{1} - \left( {Z_{2} \;\theta _{2}
} \right)\;r_{2}}  \right]}}}} \right. \kern-\nulldelimiterspace} {{\left[
{\left( {Z_{1} \;\theta _{1}}  \right)\;r_{1} - \left( {Z_{2} \;\theta _{2}
} \right)\;r_{2}}  \right]}}}^{{\kern 1pt}} {\rm ,}
\end{equation}

\noindent
where $r_{1} = c_{12} R{\kern 1pt} _{2} - c_{22} R{\kern 1pt} _{1} $;$r_{2}
= c_{11} R{\kern 1pt} _{2} - c_{21} R{\kern 1pt} _{1} $; $\theta _{j} \equiv
\rho _{j} \,d_{j} $ is the surface density (its dimensionality ${\left[
{\theta}  \right]} = g / cm^{2}$) of the $j$-th reference. Formula (\ref{eq10})
involves only the radiography data $R{\kern 1pt} _{1,2} $ and the
calibration data for measurements with two reference samples with known
$\left( {Z_{j} ;\,\rho _{j}}  \right)$ and fixed thickness $d_{j} $. The
calibration data are represented by the matrix $c_{i\,j} = R\,(E_{i} ;Z_{j}
,\theta _{j} )$, where $i,j = 1,2$. In calibration, one should account for
the solubility conditions of system (\ref{eq8}), which are, in fact, conditions on
the monitoring feasibility. Hence, it follows that the choice of the
radiation range and the references should comply with the requirements $\det
c_{ij} = c_{11} c_{22} - c_{12} c_{21} \ne 0$; $Z_{1} \ne Z_{2} $. There are
no limitations imposed on $\theta _{1,2} $, i.e., there is no limiting
relationship between density and thickness of the samples.

Let us present equation (\ref{eq10}) in the following more convenient form:

\begin{equation}
\label{eq11}
Z_{{\rm e}{\rm f}{\rm f}} = {\left[ {{{\left( {k_{1} {\kern 1pt} X + k_{2}}
\right)} \mathord{\left/ {\vphantom {{\left( {k_{1} {\kern 1pt} X + k_{2}}
\right)} {\left( {k_{3} {\kern 1pt} X + k_{4}}  \right)}}} \right.
\kern-\nulldelimiterspace} {\left( {k_{3} {\kern 1pt} X + k_{4}}  \right)}}}
\right]}^{{\kern 1pt} {{1} \mathord{\left/ {\vphantom {{1} {p}}} \right.
\kern-\nulldelimiterspace} {p}}}\quad {\rm ,}
\end{equation}

\noindent
where the new constants $k_{1} ,k_{2} ,k_{3} ,k_{4} $ are related to the old
constants from (\ref{eq10}) by certain relationships. As the formulations appear to
be rather bulky, we do not write them down explicitly. The dependence (\ref{eq11})
is a fraction-rational, non-linear, though monotonic, function. Expression
(\ref{eq11}) will not be changed under a uniform scaling transformation $k_{1}
,k_{2} ,k_{3} ,k_{4} \to Mk_{1} ,Mk_{2} ,Mk_{3} ,Mk_{4} $ ($M = {\rm c}{\rm
o}{\rm n}{\rm s}{\rm t}$). Therefore, only three of the introduced constants
are independent. They are determined by calibration using two samples of
given composition, i.e., by the matrix $c_{ij} $. However, there could be
another approach to determination of $k_{1} ,k_{2} ,k_{3} ,k_{4} $. We use,
from the beginning, the functional relationship (\ref{eq11}). The fraction-rational
function is unambiguously reconstructed from three reference points. Having
carried out the measurements for three reference materials with pre-set
values of $Z_{1} ,Z_{2} ,Z_{3} $ (note that the density and thickness of the
samples are completely arbitrary), we obtain three relationships of the form
$Z_{j}^{p} = f_{j} \left( {X_{j} ;k_{1} ,k_{2} ,k_{3} ,k_{4}}  \right)$. The
constants are then readily determined from these relationships:

\[
k_{1} = \,Z_{1}^{p} Z_{2}^{p} \left( {X_{1} - X_{2}}  \right) - \;Z_{1}^{p}
Z_{3}^{p} \left( {X_{1} - X_{3}}  \right) + Z_{2}^{p} Z_{3}^{p} \left(
{X_{2} - X_{3}}  \right)\;{\rm ;}
\]

\[
k_{2} = X_{1} X_{2} \,Z_{3}^{p} \left( {Z_{1}^{p} - Z_{2}^{p}}  \right) -
X_{1} X_{3} \,Z_{2}^{p} \left( {Z_{1}^{p} - Z_{3}^{p}}  \right) + X_{2}
X_{3} \,Z_{1}^{p} \left( {Z_{2}^{p} - Z_{3}^{p}}  \right)\;{\rm ;}
\]

\[
k_{3} = X_{1} \left( {Z_{2}^{p} - Z_{3}^{p}}  \right) - \;X_{2} \left(
{Z_{1}^{p} - Z_{3}^{p}}  \right) + X_{3} \left( {Z_{1}^{p} - Z_{2}^{p}}
\right)\;{\rm ;}
\]

\begin{equation}
\label{eq12}
k_{4} = X_{1} X_{2} \,\left( {Z_{1}^{p} - Z_{2}^{p}}  \right) - X_{1} X_{3}
\,\left( {Z_{1}^{p} - Z_{3}^{p}}  \right) + X_{2} X_{3} \,\left( {Z_{2}^{p}
- Z_{3}^{p}}  \right)\;{\rm ,}
\end{equation}

\noindent
where $X_{j} = {{R_{1j}}  \mathord{\left/ {\vphantom {{R_{1j}}  {R_{2j}} }}
\right. \kern-\nulldelimiterspace} {R_{2j}} }$ is the relative reflex (\ref{eq9})
for radiography of the $j$-th reference ($j = 1,2,3$) at fixed energies of
2-radiography. Unlike the case of $c_{ij} $-calibration for expression (\ref{eq10}),
three, and not two, reference samples are to be used here. This is related
to the fact that density and size (dimensions) of the reference samples are
not fixed. Obviously, such a calibration procedure is more convenient for
experiments and practical applications.

It is convenient to consider parameter $p$ in Eq. (\ref{eq11}) as fixed. Earlier, it
was noted that $p = 3$ for energies in the photo effect range, and $p = 1$
for the pair formation effect. At the same time, $p$ can be considered as
one more undetermined constant, especially, if it is not known beforehand
which absorption mechanisms prevail, or if the energies are used at which
all these mechanisms are essential. Than the value of $p$ can be determined
numerically by one of the approximation methods. It can minimize in the best
way the set of mean square deviations $\Delta _{n} \left( {p} \right) =
Z_{n}^{p} - {{{\left[ {k_{1} \left( {p} \right)X_{n} + k_{2} \left( {p}
\right)} \right]}} \mathord{\left/ {\vphantom {{{\left[ {k_{1} \left( {p}
\right)X_{n} + k_{2} \left( {p} \right)} \right]}} {{\left[ {k_{3} \left(
{p} \right)X_{n} + k_{4} \left( {p} \right)} \right]}}}} \right.
\kern-\nulldelimiterspace} {{\left[ {k_{3} \left( {p} \right)X_{n} + k_{4}
\left( {p} \right)} \right]}}}$ for a certain basis of calibration
measurements for substances with known atomic number values $Z_{n} $ ($n =
1,2,\ldots $). It should be noted that in the same way it is possible to
substantially increase the accuracy of effective atomic number determination
in a class of objects with close $Z_{{\rm e}{\rm f}{\rm f}} $values, for
instance, to increase the accuracy of distinction between organic
substances, etc.

\begin{center}
\textbf{4. Discussion and experiment data}
\end{center}

The proposed method of ``direct'' reconstruction of $Z_{{\rm e}{\rm f}{\rm
f}} $ is free from the above-discussed disadvantages related to the
monitoring being dependent upon the chosen basis. The ``direct'' method
ensures up to 95\% accuracy for $Z_{{\rm e}{\rm f}{\rm f}} $ determination.
In fact, using expressions (\ref{eq6})-(\ref{eq12}) to determine the relative sensitivity
$S_{Z} $ , we obtain an estimate

\begin{equation}
\label{eq13}
S_{Z} = \left( {{{\Delta Z} \mathord{\left/ {\vphantom {{\Delta Z} {Z}}}
\right. \kern-\nulldelimiterspace} {Z}}} \right) \propto 2\,\left( {{{\Delta
d} \mathord{\left/ {\vphantom {{\Delta d} {d}}} \right.
\kern-\nulldelimiterspace} {d}}} \right)\quad {\rm ,}
\end{equation}

\noindent
where $\Delta Z$ is the minimum detectable change in the effective atomic
number, and $\Delta d$ are the smallest detectable variations of the object
thickness. Sensitivity with respect to defect detection $S_{d} = \left(
{{{\Delta d} \mathord{\left/ {\vphantom {{\Delta d} {d}}} \right.
\kern-\nulldelimiterspace} {d}}} \right)$ for multi-energy radiography
corresponds to the sensitivity of conventional radiography (separately for
each of the assembly detectors) and is normally of the order of several
percent. The numerical factor 2 in the expression (\ref{eq13}) is related to the
two-energy nature of the monitoring (i.e., sensitivity is assumed to be
equal for all detectors, $S_{d,1} = S_{d,2} $). Consequently, $S_{Z} \propto
5\% $, if $S_{d} \propto 2.5\% $. This requires spatial resolution of
$4\;$pl/mm .

The relationship (\ref{eq13}) determines the accuracy of the proposed method. It can
be seen that 95\% reconstruction of the effective atomic number can be
achieved for an unknown material. Below, this conclusion is confirmed by
experimental data. As noted before, the reason for such dramatic increase in
accuracy is related to the direct reconstruction of from radiography
measurements without using a fitting procedure by choosing the ``basic''
materials. A formal representation (``replacement'') of an arbitrary
material by superposition of two specified materials with known absorption
coefficients leads to large errors in determination of $Z_{{\rm e}{\rm
f}{\rm f}} $. The existing materials are too numerous for such rough
approximation for absorption of electromagnetic radiation. On the contrary,
the direct effective atomic number using only the relative radiographic
reflex (\ref{eq9}) for the inspected material allows us to avoid this mistake. It is
of an order of $\left( {\Delta Z} \right)_{{\rm i}{\rm n}{\rm d}{\rm i}{\rm
r}{\rm e}{\rm c}{\rm t}} \propto {{Z} \mathord{\left/ {\vphantom {{Z} {M}}}
\right. \kern-\nulldelimiterspace} {M}}$ ($M$ is the order of multiplicity),
i.e., for indirect reconstruction of $Z_{{\rm e}{\rm f}{\rm f}} $ in
two-energy radiography the errors can reach 50\% . Many industrial
radiographic installations using indirect methods have such accuracy,
$\Delta Z = \pm 0.5Z_{{\rm e}{\rm f}{\rm f}} $. This is quite sufficient to
discern organics from inorganics, e.g., to see the difference between heavy
alloys, $Z_{{\rm e}{\rm f}{\rm f}} = 26\pm 13$, and plastics $Z_{{\rm e}{\rm
f}{\rm f}} = 6\pm 3$, but is clearly not enough for more accurate
measurements in many applications. The direct method of $Z_{{\rm e}{\rm
f}{\rm f}} $ reconstruction opens here many new possibilities.

In general, unlike the ``synthesis'' of $Z_{{\rm e}{\rm f}{\rm f}} $, the
direct method is based upon ``analysis'' of the atomic composition.
Moreover, in the new approach it is possible to limit oneself to the use of
just the 2-radiography. This is important, because passing over to
radiography with higher multiplicity is a technically difficult task. To
verify the theory, we compared the obtained theoretical dependence (\ref{eq11})-(\ref{eq12})
with known experimental data on gamma-radiation absorption in a large range
of various materials, starting from carbon ($Z = 6$) and ending with uranium
($Z = 92$). These results are shown in Fig.~2-4. In constructing the
theoretical curves, three points were chosen as reference ones (defining a
fraction-rational function), which corresponded to materials with large,
small and intermediate $Z_{{\rm e}{\rm f}{\rm f}} $ values. Three
characteristic regions of the energy spectrum were considered -- i.e., low,
middle and high energies, corresponding, respectively, to the photo effect,
the Compton effect and the pair production effect. The data presented are in
a very good agreement with theory.

The choice of absorption mechanisms depends upon the radiation energies
used. The gamma-quanta energy range is mainly determined by the experiment
conditions of character of applications. E.g., in security and customs
inspection of luggage, medium energies from several tens to several hundreds
keV are used. This corresponds to the region of combined effects of photo
absorption and Compton scattering of gamma-quanta in the substance (Figs.
2-3). Discarding any of these mechanisms in the medium energy range can lead
to substantial errors in determination of the effective atomic number. Thus,
in Fig.~3 a theoretical approximation neglecting the photo effect gives much
worse results. For inspection of large objects, such as trucks or
containers, accelerator energies of several MeV are needed. In this case,
for determination of the effective atomic number one should use a model
where the pair formation effect and Compton effect are predominant It should
be stressed that the proposed method can be used with any choice of the
radiation energies. For the data presented in Figs. 2-4, we used only
several fixed radiation energy values. We have checked that with any other
choice of these energies the principal law (\ref{eq11}) remains valid. For the
experimental data used, deviations from the theoretical dependence did not
exceed 5-10\% , which corresponds to 90-95\% accuracy in $Z_{{\rm e}{\rm
f}{\rm f}} $ determination.

Detailed analysis of the optimal design for 2-radiography,
completion of direct experiments, reconstruction of the atomic
composition of different objects and complex materials, micro- and
macro-structures, etc. are the subjects of our further studies.
The approach presented in this paper allows the most complete
extraction of the information on physical properties of the
material studied by multi-energy radiography. The number of
parameters reconstructed corresponds to the radiography
multiplicity. (This is a general principle of multi-energeticity).
This direct analytical approach uses only absorption data, showing
up the amplitude contrast induced by ionizing radiation of photons
(as monochromatic electromagnetic waves). Together with new X-ray
analysis methods, including those using synchrotron (coherent)
radiation exposing the phase contrast, as well as multi-energy
holography \cite{Wilkins:1996}, \cite{Gog:1996},
\cite{Cloetens:1999}, micro-focusing and computer tomography
\cite{Van:2000}, \cite{See:2000}, etc., the proposed approach to
monitoring of materials structure widely broadens the field of NDT
possibilities. This direction is similar to new directions in
studies of semiconductor and other films. Also, certain prospects
are opened for application of NDT in development of various
microstructures, as well as in studies of the distribution profile
of implanted nano-clusters of alien atoms in the ``host'' crystal
lattice.

\begin{center}
\textbf{5. Summary}
\end{center}

Thus, a direct approach has been proposed to reconstruction of the atomic
structure of materials by means of multi-energy radiography. The general
expressions (\ref{eq10})-(\ref{eq12}) obtained for the effective atomic number in
2-monitoring, are closed and convenient for development of computer data
processing algorithms. These formulas can be also used for reconstruction of
the structure of mixtures, composites, multi-component objects, micro- and
macro-structures, systems with variable chemical composition, etc. The
validity of a universal law (fraction-rational dependence with calibration
over three reference points) in the atomic number radiography is
experimentally confirmed in different energy ranges of X-ray and
gamma-radiation. An essential feature of multi-energy radiography is direct
extraction of additional information on physico-chemical structure of the
object under study. This opens new possibilities in materials science,
non-destructive testing and other applied fields of science and technology.

\bigskip

\begin{center}
\textbf{Acknowledgements}
\end{center}

The research described in this publication was made possible in part by
Award No. UE2-2484-KH-02 of the U.S. Civilian Research \& Development
Foundation for the Independent States of the Former Soviet Union (CRDF).

\newpage
\textbf{Table 1.} \textbf{Effective atomic number of various substances with
respect to the photo effect,} $Z_{{\rm p}{\rm h}{\rm o}{\rm t}{\rm o}}
$\textbf{, pair formation effect,} $Z_{{\rm p}{\rm a}{\rm i}{\rm r}{\rm s}}
$\textbf{, and Compton scattering,} $Z_{{\rm c}{\rm o}{\rm m}{\rm p}{\rm t}}
$\textbf{.}

\newcommand{\PreserveBackslash}[1]{\let\temp=\\#1\let\\=\temp}
\let\PBS=\PreserveBackslash
\begin{table}
\begin{tabular}
{|p{130pt}|p{108pt}|p{30pt}|p{30pt}|p{32pt}|} \hline Material&
Chemical formula& $Z_{photo}$ & $Z_{pairs}$ & $Z_{compt}$
 \\
\hline Inorganic substances& & & &
 \\
\hline Stainless steel& Fe 66\% ; Cr 10\% ; Ni 16\% ; Ti 8\% &
26.57& 27.58& 27.80 \\ \hline Black steel& Fe 92\% ; C 8\% &
25.97& 25.94& 25.76 \\ \hline Calcium phosphate; bone tissue
(med.)& Ca(PO$_{{\rm 4}}$)$_{{\rm 2}}$& 17.38& 16.11& 11.05 \\
\hline Table salt& NaCl& 15.66& 15.21& 14.62 \\ \hline Quartz
glass; sand& SiO$_{{\rm 2}}$& 12.30& 11.63& 10.80 \\ \hline
Aluminum and light alloys& Al$_{{\rm 2}}$O$_{{\rm 3}}$& 11.70&
11.23& 10.65 \\ \hline Glass& Na$_{{\rm 2}}$SiO$_{{\rm 3}}$&
11.49& 11.02& 10.51 \\ \hline Water& H$_{{\rm 2}}$O& 7.98& 7.89&
7.22 \\ \hline Air& mixture O$_{{\rm 2}}$; N$_{{\rm 2}}$ etc.&
7.6& 7.4& 6.9 \\ \hline Organic substances& & & &
 \\
\hline Polyvinyl chloride& (C$_{{\rm 2}}$H$_{{\rm 3}}$Cl)$_{{\rm
n}}$& 15.85& 14.80& 11.97 \\ \hline Soft tissue (med.)&
CNO-organics; H$_{{\rm 2}}$O~90\% & 7.8& 7.2& 6.8 \\ \hline
Glucose& C$_{{\rm 6}}$H$_{{\rm 1}{\rm 2}}$O$_{{\rm 6}}$& 7.37&
7.22& 6.73 \\ \hline Saccharose& C$_{{\rm 1}{\rm 2}}$H$_{{\rm
2}{\rm 2}}$O$_{{\rm 1}{\rm 1}}$& 7.38& 7.18& 6.71 \\ \hline
Cellulose (wood, fabrics)& (C$_{{\rm 6}}$H$_{{\rm 1}{\rm
0}}$O$_{{\rm 5}}$)$_{{\rm n}}$& 7.31& 7.14& 6.68 \\ \hline Organic
glass& (C$_{{\rm 5}}$H$_{{\rm 8}}$O$_{{\rm 2}}$)$_{{\rm n}}$&
6.96& 6.76& 6.24 \\ \hline Polyamide (nylon)& (C$_{{\rm
6}}$H$_{{\rm 1}{\rm 1}}$NO$_{{\rm 2}}$)$_{{\rm n}}$& 6.85& 6.70&
6.18 \\ \hline Polystyrene& (C$_{{\rm 8}}$H$_{{\rm 9}}$)$_{{\rm
n}}$& 5.95& 5.92& 5.57 \\ \hline Polyethylene (plastics)&
(C$_{{\rm 2}}$H$_{{\rm 4}}$)$_{{\rm n}}$& 5.94& 5.86& 5.29 \\
\hline
\end{tabular}
\end{table}

\newpage
\textbf{Figure captions}

\bigskip

FIG.~1. General scheme of two-energy radiography with
reconstruction of the effective atomic number of the material.
Synthesis consists in mixing of the fitting basic elements with
\textbf{\textit{L}} -- \textbf{"light"}, \textbf{\textit{M}} --
\textbf{"middle"} and \textbf{\textit{H}} -- \textbf{"heavy"}
atomic mass. Analysis is unambiguous reconstruction of $Z_{eff}$.
``Black-and-white'' synthesis corresponds to the two-energy
radiography, ``three-color'' scheme (\textbf{\textit{R}} --
\textbf{"red"}, \textbf{\textit{G}} -- \textbf{green}, and
\textbf{\textit{B}} -- \textbf{"blue"}) corresponds to
3-radiography, etc. For the direct method (analysis) proposed in
this work, 2-radiography is sufficient.

FIG.~2. Dependence of the effective atomic number upon
2-radiography reflexes in the regions of photo effect and Compton
scattering. Theoretical dependence (solid line) and experimental
points for materials of known composition are indicated. Geometry
(dimensions) of samples is arbitrary.

FIG.~3. Dependence of the effective atomic number upon
2-radiography reflexes in the region of intermediate ionizing
radiation energies. Theoretical curves are presented modeling the
predominant role of one of two possible absorption channels. The
best agreement with experimental data is obtained when a mixed
absorption mechanism (photo effect/Compton effect) is chosen.

FIG.~4. Dependence of the effective atomic number upon
2-radiography reflexes in the region of Compton scattering and the
pair formation effect. When the detected energy ranges are moved
apart, sensitivity of the method is not worsened.

\newpage
\begin{figure}
\centering
\includegraphics*[scale=0.70]{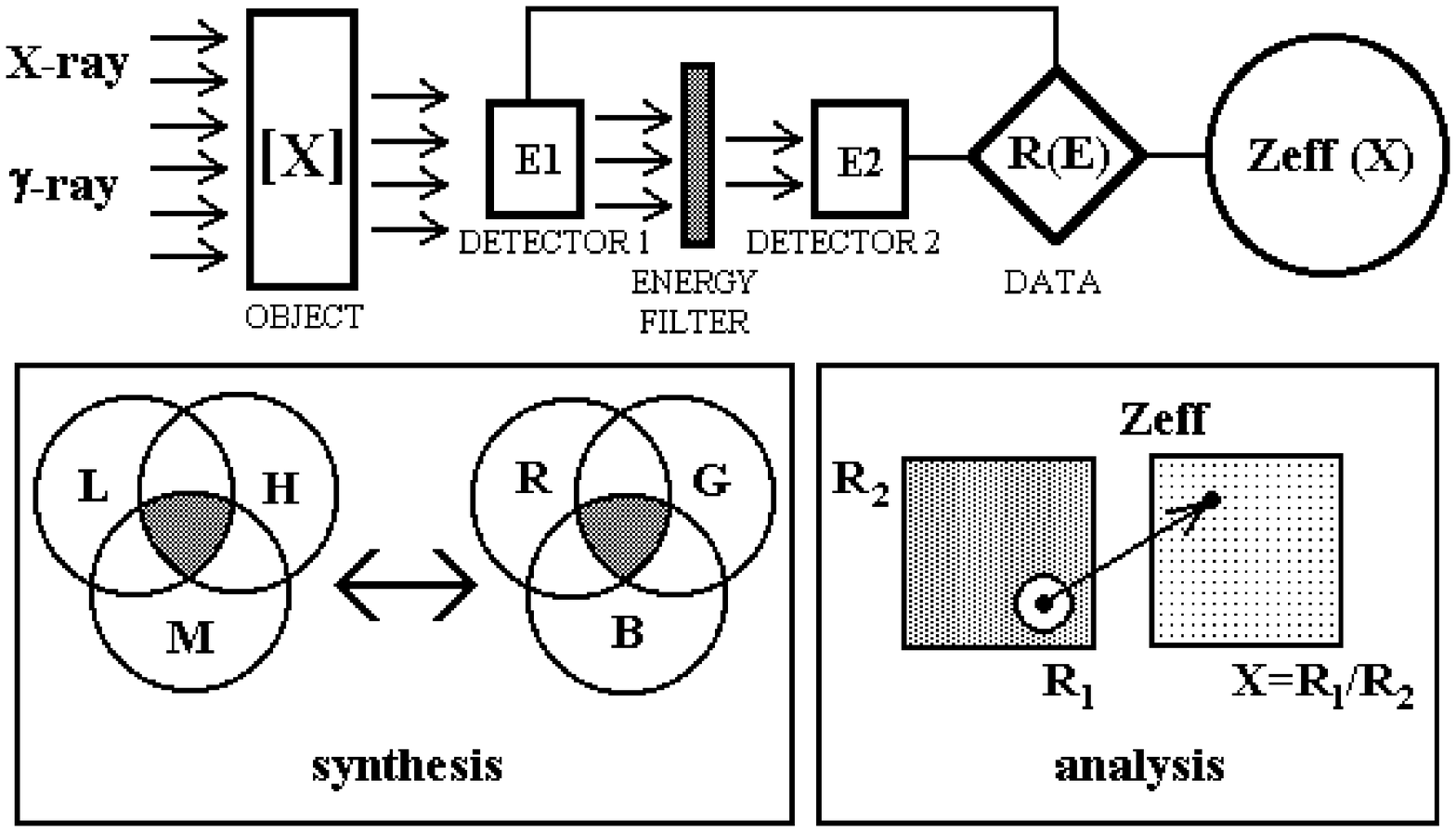}
\caption{ }
\end{figure}

\newpage
\begin{figure}
\centering
\includegraphics[width=5.17in,height=6.41in]{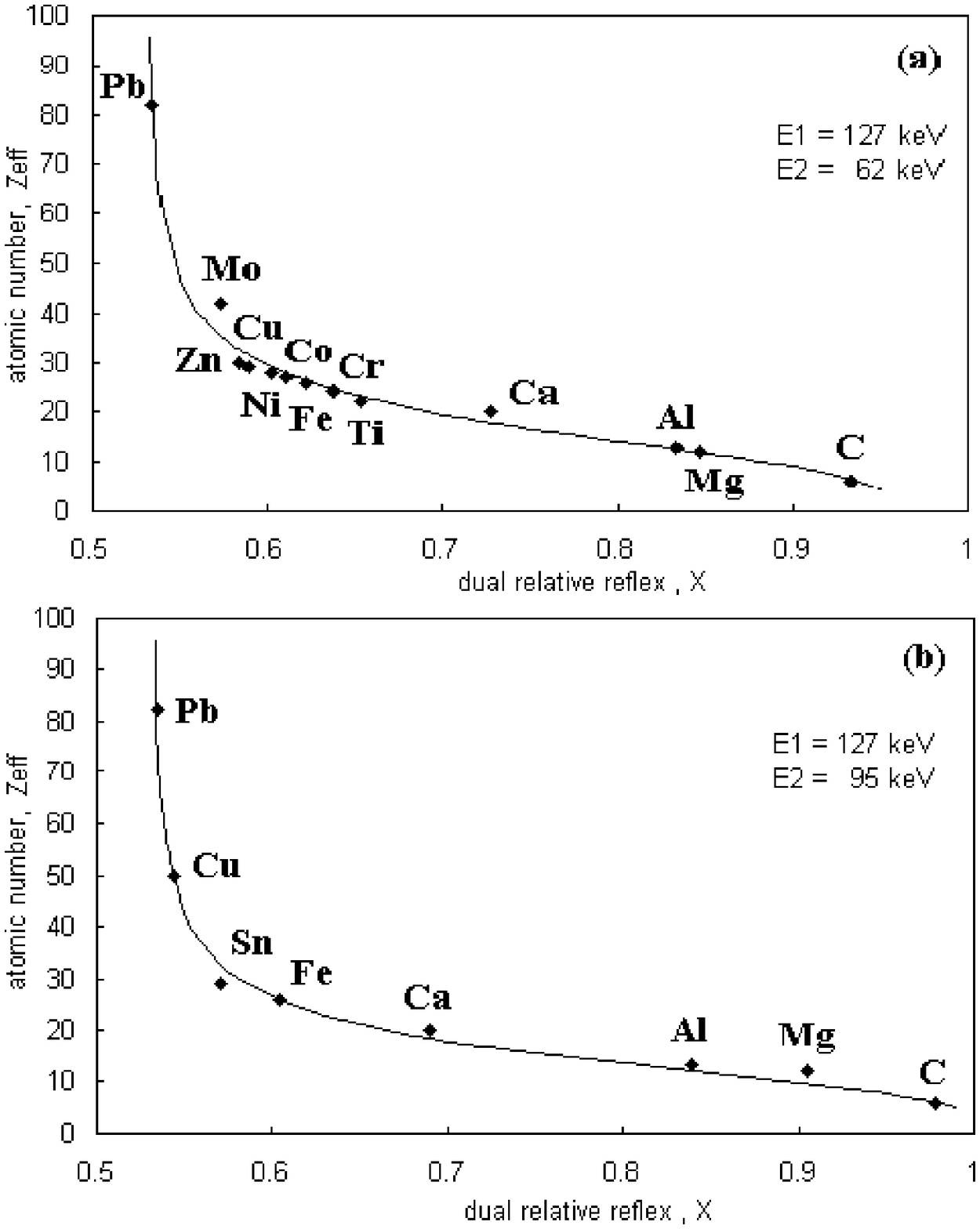}
\caption{ }
\end{figure}

\newpage
\begin{figure}
\centering
\includegraphics*[scale=0.70]{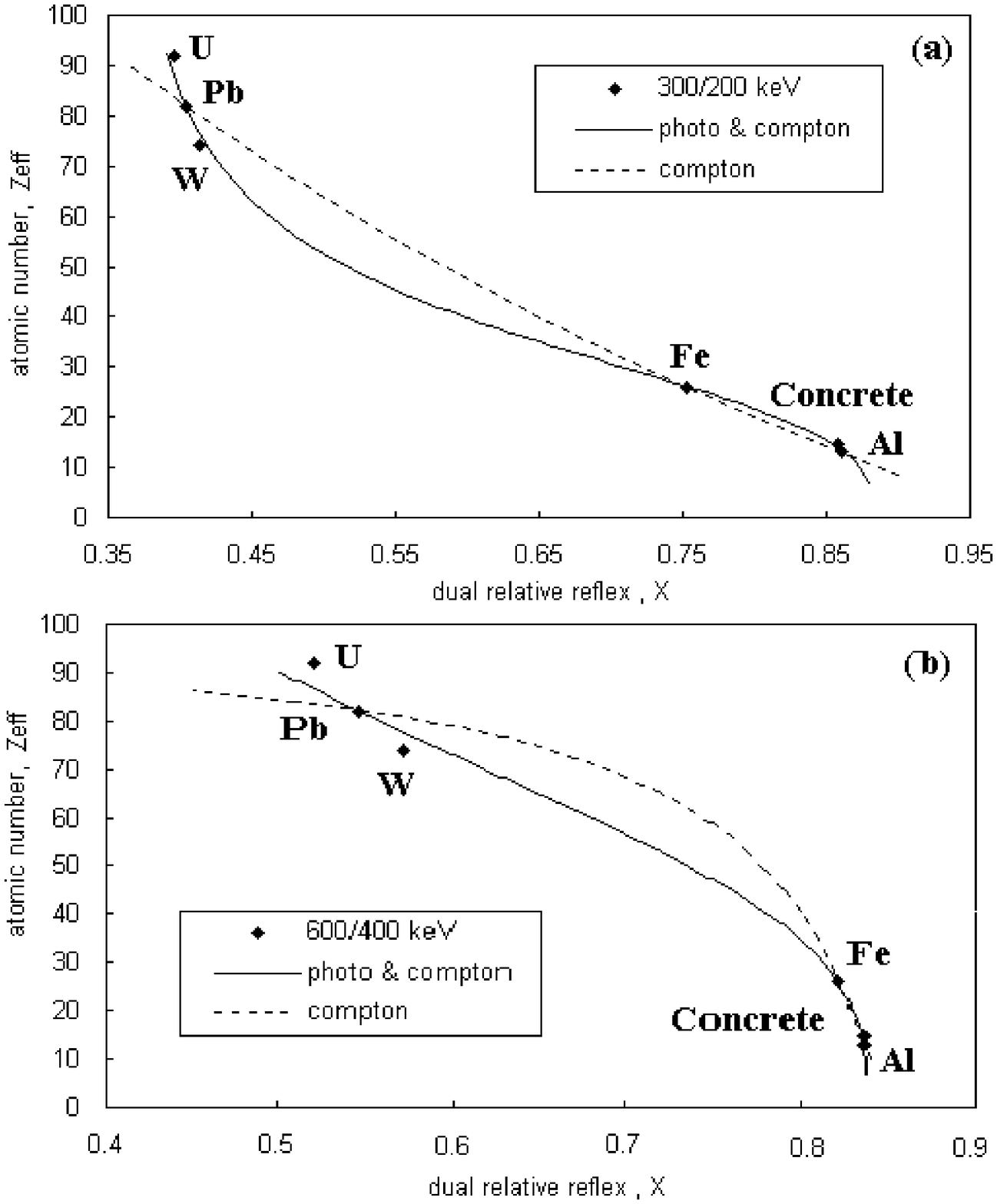}
\caption{ }
\end{figure}

\newpage
\begin{figure}
\centering
\includegraphics[width=5.40in,height=6.55in]{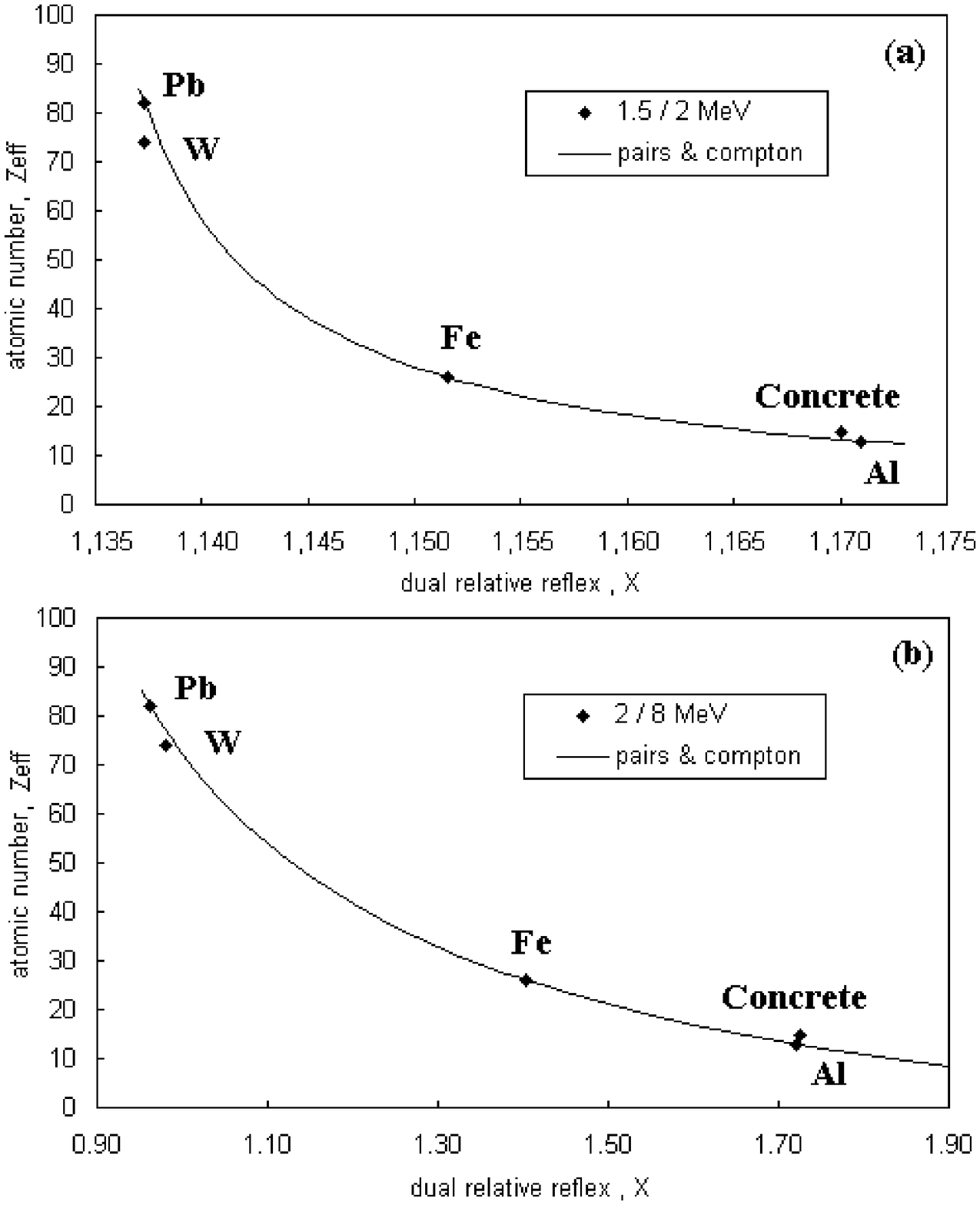}
\caption{ }
\end{figure}

\end{document}